\documentclass[useAMS,usenatbib,usedcolumn,usegraphicx]{mn2e}

\def\en{{$n$}\ }

\def\fexvii{{\rm Fe~\sc xvii}\ }
\def\fexviii{{\rm Fe~\sc xviii}\ }
\def\fexix{{\rm Fe~\sc xix}\ }

\usepackage[dvips]{graphics}
\usepackage[dvipsnames,usenames]{color}
\usepackage{lscape}

\newcommand{\be}{\begin{equation}}
\newcommand{\ee}{\end{equation}}

\title{Interface of Equation-of-State, Atomic Data and
Opacities in the Solar Problem} 
\author
[Anil K.\ Pradhan]
       {Anil K. Pradhan$^{1,2}$\\
       $^1$ Department of Astronomy, $^2$ Chemical Physics Program,
 The Ohio State University, Columbus, OH 43210, USA.}
\date{Accepted  xxxxxx 
      Received xxxxxx;
      in original form xxxxxx}

\pagerange{\pageref{firstpage}--\pageref{lastpage}}
\pubyear{2022}

\def\LaTeX{L\kern-.36em\raise.3ex\hbox{a}\kern-.15em
    T\kern-.1667em\lower.7ex\hbox{E}\kern-.125emX}

\begin{document}

\maketitle

\label{firstpage}

\begin{abstract}

 Convergence of the Rosseland Mean Opacity (RMO) is investigated with
respect to the equation-of-state (EOS) and the number of atomic levels
of iron ions prevalent at the solar radiative/convection
boundary. The "chemical picture"
Mihalas-Hummer-D\"{a}ppen MHD-EOS, and its variant QMHD-EOS, are studied at
two representative temperature-density sets at the base of the
convection zone (BCZ) and the Sandia Z experiment: 
$(2 \times 10^6K, \ 10^{23}/cc)$ and $(2.11 \times
10^6K, \ 3.16 \times 10^{22}/cc)$, respectively. 
It is found that whereas the new atomic datasets from accurate R-matrix
calculations for opacities (RMOP) are vastly
overcomplete, involving hundreds to over a thousand levels of
each of the three Fe ions considered --- \fexvii, \fexviii,
\fexix --- the EOS constrains contributions to RMOs by relatively fewer levels. 
The RMOP iron opacity spectrum is quite different from the Opacity
Project distorted wave model and shows considerably more 
plasma broadening effects. This work points to possible
improvements needed in the EOS for
opacities in high-energy-density (HED) plasma sources. 
\end{abstract}


\section{Introduction}
As a fundamental quantity in light-matter interaction opacity plays a key 
role in astrophysics, such as stellar interiors, helioseismology, and 
asteroseimology,
elemental abundance determination, host-star and exoplanetary fluxes, 
etc. (\cite{jcd,basu15,a09,c19,b23a}. 
In addition, radiation transport models of inertial 
plasma fusion devices requires accurate opacities (\cite{b15,p18}.
In particular, the outstanding uncertainty in the solar
chemical composition
affects elemental calibration of all astronomical sources. Attempts to employ
advances in helioseismology and abundances are an active area of basic
research (\cite{ba08,b22}, 
but require enhanced solar opacities by about 10\%.  That, in turn,
depends on two elements, oxygen and iron, that determine about half of
the solar opacity at BCZ. However, a downward revision of oxygen
abundance by up to 20-40\% from earlier solar composition
is a major part of the "solar problem" (\cite{aag21,pie23,li23,b23b}. Since about 90\% of oxygen is 
either fully ionized or H-like at BCZ, its
absorption coefficient is small and 
unlikely to change from current atomic
calculations, enhanced iron opacity might countenance lower
solar abundances (\cite{b15}.

Opacity computations depend on atomic data on the one hand and the
plasma EOS on the other (\cite{op,symp,p1}. 
Voluminous amounts of data are needed
for all photon absorption and scattering processes in order to ensure
completeness. Recently, accurate and extensive calculations of atomic data for iron
ions of importance under BCZ conditions have been carried out using the
R-matrix method (\cite{p1,p2,p3,p4}. 
However, the EOS determines how and to what extent the
atomic data contribute to monochromatic and mean opacities at a given
temperature and density. The Planck and Rosseland Mean Opacity (PMO and
RMO respectively) are defined as

\be \kappa_P B(T) = \int \kappa_\nu B_\nu d\nu, \ee

\begin{equation}
 \frac {1}{\kappa_R} = \frac{\int_0^\infty g(u) \kappa_\nu^{-1}
du}{\int_0^\infty g(u) du} \ \ \ \hbox{\rm ; }\ \ \ g(u) = u^4 e^{-u}
(1 - e^{-u})^{-2},
\label{eq:RMO}
\end{equation}

where $g(u) = dB_\nu/dT$ is the derivative of the Planck weighting
function
\be B_\nu(T) = \frac{(2h\nu^3/c^2)}{e^{h\nu/kT}-1} \ee, 

and $\kappa_\nu$ is the monochromatic opacity.
Atomic processes and contributions to opacity are from bound-bound
($bb$), bound-free ($bf$), free-free ($ff$), and
photon scattering ($sc$) as
\begin{eqnarray}
 \kappa_{ijk}(\nu) & = & \sum_k a_k \sum_j x_j \sum_{i,i'}
\ [\kappa_{bb(}(i,i';\nu) \\
& + & \kappa_{bf}(i,\epsilon i';\nu)  +  \kappa_{ff} (\epsilon i, \epsilon' i';
\nu) + \kappa_{sc} (\nu)]\ ,
\label{eq:k}
\end{eqnarray}

where $a_k$ is the abundance of element $k$, $x_j$ the $j$ ionization
fraction, $i$ and $i'$ are the initial bound and final bound/continuum
states of the atomic species, and $\epsilon$ represents the electron
energy in the continuum. Whereas the $ff$ and $sc$ contributions are
small, the opacity is primarily governed by $bb$ and $bf$ atomic data
that need to be computed for all atomic species. Existing opacity
models generally employ
the relatively simple distorted wave (DW) approximation based on
atomic structure codes, but
higher accuracy requires considerable effort. 

Originally, the Opacity Project (\cite{op} (hereafter OP)
envisaged using the poweful and highly accurate R-matrix 
method for improved accuracy. But that turned out to be intractable owing to
computational constraints, and also required theoretical developments related to
relativistic fine structure and plasma broadening effects. Therefore,
the OP opacities were finally computed using similar atomic physics
as other existing opacity models, mainly based on the simpler 
distorted wave (DW) approximation (\cite{opcd}, 
and later archived in the online database OPserver (\cite{ops}.
However, following several developments since then
renewed R-matrix calculations can now be carried out, as discussed below. 

\section{Theoretical framework}

Recently, with several improvements in
the extended R-matrix and opacity codes large-scale data have been computed 
for Fe ions \fexvii, \fexviii and \fexix, which determine 
over 80\% of iron opacity near BCZ conditions (\cite{p1,p2,p3,p4}.
The R-matrix (RM) framework and comparison with existing opacity models
based on atomic structure codes and the distorted wave (DW) approximation, 
and associated physical effects,
are described in detail. The primary difference between the RM and DW approximations is the treatment of bound-free opacity which is dominated by autoionizing resonances that are included in an {\it ab initio} manner in RM calculations, but treated perturbatively
 as bound-bound transitions in the DW method.
Plasma broadening effects are very important, but 
manifest themselves quite differently in the 
two methods. Resonances in
RM photoionization cross sections are broadened far more than lines
as function of temperature and density since
autoionization widths, shapes and heights
 are considered explicitly
(\cite{p3}. Also, the intrinsically asymmetric features of the large
Seaton photoexcitation-of-core (PEC) resonances in bound-free cross sections
are preserved in RM calculations. The unverified assertion that
RM and DW opacities are equivalent is incorrect owing to basic
physical effects (\cite{d21}. On the
contrary, the RM method is based on the coupled channel approximation
that gives rise to autoionizing resonances, and has historically superseded 
the DW method which neglects
channel coupling. RM calculations for all relevant 
atomic processes are generally much
more accurate than the DW, as for example in the work carried out under the
Iron Project, including relativistic effects in the Breit-Pauli R-matrix
(BPRM) approximation (\cite{ip} that is also employed in the present work
(\cite{p2}. 

 The interface of atomic data with EOS parameters is implemented 
through the MHD-EOS (\cite{mhd}, formulated in the "chemical picture" as
designed for OP work. It is based on the concept of {\it
 occupation probability} $w$ of an atomic level being populated in a
plasma environment, characterized by a temperature-density
(hereafter T-D) related to Boltzmann-Saha equations.
 The level population is then given as

 \be N_{ij} = \frac{N_j  g_{ij} w_{ij} e^{-E_{ij}/kT}}{U_j}, \ee

 where $w_{ij}$ are the occupation probabilities of levels $i$ in
  ionization state $j$, and $U_j$ is the atomic internal partition
function. The occupation probabilities do not have a sharp
  cut-off, but approach zero for
  high-\en as they are "dissolved" due to plasma interactions.
 The partition function is re-defined as

 \be U_j = \sum_i g_{ij} w_{ij} e^{(-E_{ij}/kT)}. \ee

$E_{ij}$ is the excitation energy of level $i$, $g_{ij}$ its statistical
weight, and $T$ the temperature. The $w_{ij}$ are determined upon
free-energy minimization in the plasma at a given T-D. However, the
original MHD-EOS was found to yield $w$-values that were
unrealistically low by up to several orders of magnitude. 
An improved treatment of microfield distribution and
plasma correlations was developed, leading to the
so-called QMHD-EOS (\cite{qmhd} and employed for
subsequent OP calculations and results (\cite{opcd,ops}.

\section{Opacity computations}

The new RMOP data are interfaced
with the (Q)MHD-EOS to obtain opacities.
Computed RM atomic data for
$bb$ oscillator strengths and $bf$ photoionization cross sections of all
levels up to \en(SLJ) = 10 yields
datasets for 454 levels for \fexvii, 1174 levels for \fexviii and 1626
for \fexix (\cite{p2}; some results for \fexvii were reported 
earlier (\cite{np16}. 
Monochromatic and mean opacities may then be computed 
using atomic data for {\it any number of these levels and the EOS}.

 In order to study the behavior of MHD and QMHD, we employ 
the new RMOP opacity codes (\cite{p1}, varying the number of atomic
levels for each Fe ion, and both sets
of EOS parameters at specified temperature-density pairs for a
particular ion. Monochromatic opacities are computed at the same
frequency mesh in the variable and range $0 \leq u=h\nu/kT \leq 20$, 
as in OP work (\cite{symp,ops}.
Since RMOP calculations were carried out for
the three Fe ions that comprise over 80\% of total Fe at BCZ, 
we replace their opacity spectra in OP codes (\cite{opcd} 
and recompute RMOP iron
opacities. Thus, $\sim$15\% contribution is from OP data for other Fe
ions; a table of Fe ion fractions at BCZ is given in (\cite{p1}.

 To circumvent apparently unphysical behavior of MHD-EOS at very high
densities, an ad hoc occupation probability cut-off was introduced in OP
calculations with $w(i) \geq
0.001$ (\cite{bs03}. We retain the cut-off in the new RMOP opacity codes
(\cite{p1}, since the same EOS is employed, but also tested relaxing the cut-off 
to smaller values up to $w(i) \geq 10^{-12}$. However, no significant
effect on RMOs was discernible, indicating that a more fundamental
revision of (Q)MHD-EOS might be necessary (\cite{rt}. Level
population fractions are normalized to unity, and therefore including
more levels would not necessarily affect opacities in a systematic
manner, as discussed in the next section.
unless they are modified with inclusion of 
possibly missing atomic-plasma microphysics of
individual levels and associated atomic data.

\section{Results and discussion}

\begin{figure} 
\includegraphics[width=\columnwidth,keepaspectratio]{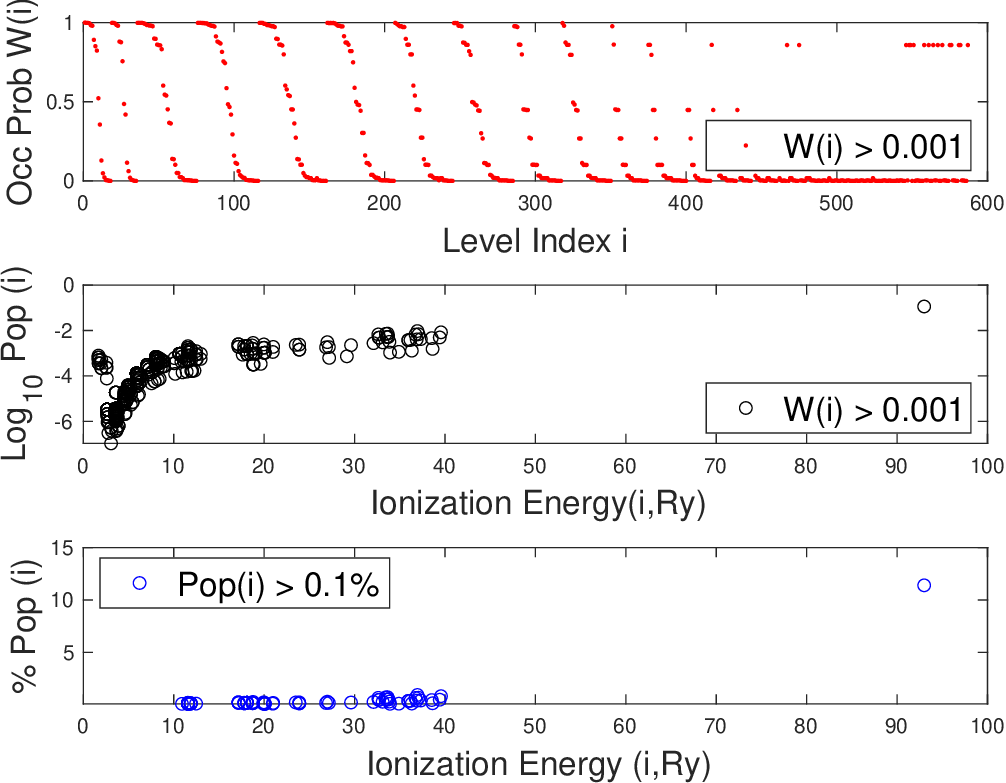}
\caption{\fexvii EOS parameters at BCZ conditions: 
occupation probabilities w(i) as function of
level index $i$ (top, red dots); $Log_{10}$ of level populations
$Pop(i)$ vs.  ionization energy
(middle, black open circles); levels with percentage $Pop(i) > 0.1$\% 
vs. ionization energy. The ground state population is 11\% and the
ionization energy is 93 Ry.
The $w(i)$ (top panel) correspond to levels $i$ computed along 
spin-orbital-parity 
SLJ$\pi$ symmetries of bound levels in RMOP computations (see text).
 \label{fig:eos}}
\end{figure}

The EOS determines the contribution to opacity and its cut-off 
from an atomic level $i$
via the occupation probability $w(i)$ depending on density and resulting
plasma microfield, and the level population $pop(i)$ via the Boltzmann
factor $exp(-E_i/kT)$ at temperature T. Fig.~\ref{fig:eos} illustrates
the behavior of the EOS parameters for \fexvii at BCZ conditions.
The new RMOP data include autoionizing resonances due to 
several hundred coupled levels, but can not be directly compared with
DW bound-free 
cross sections that neglect channel coupling and are feature-less 
(\cite{p2,p4}. 
However, a comparison of the total monochromatic opacity
spectrum can be done to illustrate differences due to plasma broadening of
resonances in the RMOP data vs. lines as in the OP DW data. 

The primary focus of this work is the interface of EOS with atomic
data. As exemplar of the detailed analysis of EOS parameters, 
Fig.~\ref{fig:eos} shows the occupation probabilities for \fexvii
at BCZ conditions (red dots, top panel) for all levels with $w(i) > 0.001$, and
corresponding level populations (black open circles, middle panel). 
Since the contribution
to RMO is limited by significant level populations $Pop(i)$, the
number of levels with $Pop(i) > 0.1$\% is found to be
much smaller, around 50 or so (blue dots, bottom panel). The reason for
the given distribution of $w(i)$ (top panel) is because the BPRM calculations
are carried out according to total angular momentum quantum number and
parity $J\pi$. Therefore, all BPRM data are produced in order of
ascending order in energy {\it within each $J\pi$ symmetry}, and
descending order due to Stark ionization and dissolution of levels 
(\cite{mhd}.

\begin{figure} 
\includegraphics[width=\columnwidth,keepaspectratio]{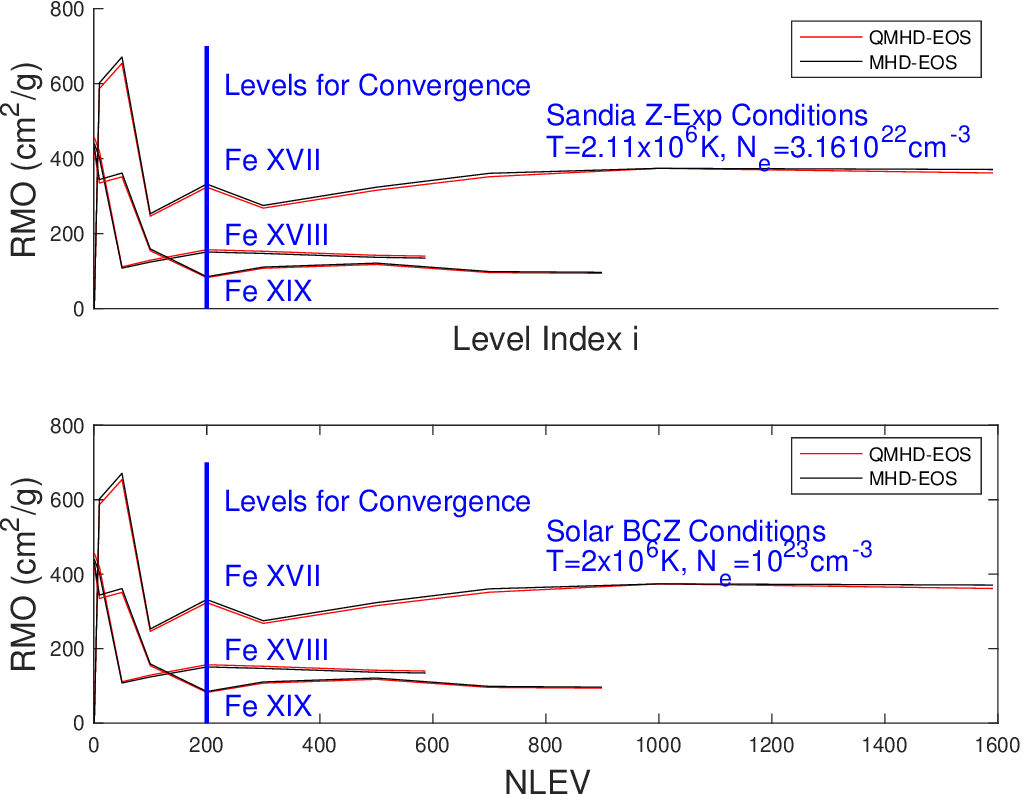}
\caption{Rosseland Mean Opacity vs. number of levels included in RMOP
opacity computations for BCZ and Sandia Z conditions. 
RMOs appear to 'converge' to constant values around NLEV
$\approx$ 200 (however, see text).
 \label{fig:rmo}}
\end{figure}

Tables 1 and 2 give sample RMOs computed at BCZ and
Sandia Z temperatures and densities respectively,
varying the number of contributing
levels NLEV for each of the three Fe ions, and both the MHD and QMHD EOS. 
Correspondingly, an illustration of RMO behavior is shown in
Fig.~\ref{fig:rmo}.
There is considerable variation in RMO values for small NLEV as
expected.  The RMOs are very high if all the population is in the ground
stae or the first few excited states, but decreasing with NLEV. But then
the RMOs approach near-constant values for NLEV $\approx$ NMAX = 200, 
for all three Fe ions and for
both the MHD and QMHD; no further significant contribution to RMOs is
made due to EOS cut-offs and saturation. 
{\it Therefore, this 'convergence' should be treated
as apparent, and would be real if and only if
the EOS is precisely determined}. The converged RMOs
should be regarded as a lower
bound, in case revisions to EOS enable contributions from more levels that are
included in the extensive RMOP atomic datasets, and the EOS+data
combination may yield higher opacities.

\begin{table}
\caption{Convergence of the Rosseland Mean Opacity (cm$^2$/g) with QMHD and MHD
equation-of-state for $T = 2 \times 10^6K, N_e \ = \ 10^{23} cc$. 
NLEV = number of bound levels in EOS calculations,
and NMAX = maximum number of bound levels in R-matrix atomic calculations.}
\vspace{0.1in}
\begin{tabular}{|ccc|cc|cc|}
\hline
 &  \fexvii & &   \fexviii  & & \fexix & \\
\hline
NLEV & QMHD & MHD & QMHD & MHD & QMHD & MHD  \\
\hline
 1 & 873.4 & 891.9 & 0.92 & 1.0 & 69.1 & 75.6 \\
 10 & 831.0 & 844.4 & 324.8 & 365.5 & 55.2 & 60.3 \\
 50 & 225.9 & 230.3 & 357.3 & 392.0 & 56.8 & 62.1 \\
 100 & 265.5 & 270.3 & 136.8 & 150.1 & 23.1 & 25.3 \\
 200 & 346.5 & 352.5 & 175.3 & 192.4 & 10.7 & 11.7 \\
 300 & 360.4 & 366.6 & 145.5 & 159.6 & 13.9 & 15.3 \\
 500 & - & - & 169.2 & 185.7 & 15.5 & 16.6 \\
\hline
 700 & - & - & 189.4 & 207.9 & 12.5 & 13.7 \\
\hline
 1000 & - & - & 197.9 & 217.2 & - & - \\
\hline
\multicolumn{7}{c} {Converged RMOs with NLEV = NMAX} \\
\hline
 587 & 352.6 & 358.7 &  - & - & - & - \\
\hline
 1591 & - & - & 196.5 & 215.6 & - & -  \\
\hline
 899 & - & - & - & - & 12.5 & 13.7  \\
\hline
\end{tabular}
\end{table}

\begin{table}
\caption{Convergence of RMOs (cm$^2$/g) with QMHD-EOS and MHD-EOS
at Sandia Z $T = 2.11 \times 10^6K, N_e \ = \ 3.16 \times 10^{22} cc$.} 
\vspace{0.1in}
\begin{tabular}{|ccc|cc|cc|}
\hline
 &  \fexvii & &   \fexviii  & & \fexix & \\
\hline
NLEV & QMHD & MHD & QMHD & MHD & QMHD & MHD  \\
\hline
 1 & 456.4 & 440.0 & 1.60 & 1.64 & 419.2 & 431.1 \\
 10 & 419.8 & 403.0 & 586.6 & 602.0 & 334.8 & 344.0 \\
 50 & 111.2 & 107.9 & 654.0 & 670.9 & 351.2 & 361.4 \\
 100 & 129.0 & 124.1 & 246.4 & 252.8 & 154.4 & 159.0 \\
 200 & 156.9 & 150.9 & 323.7 & 332.0 & 82.6 & 85.0 \\
 300 & 152.8 & 147.0 & 267.9 & 274.9 & 107.5 & 110.7 \\
 500 & 142.1 & 136.7 & 315.5  & 323.6 & 117.7 & 121.2 \\
\hline
 700 & - & - & 351.6 & 360.7 & 96.0 & 98.7 \\
\hline
 1000 & - & - & 374.0 & 374.0 & - & - \\
\hline
\multicolumn{7}{c} {Converged RMOs with NLEV = NMAX} \\
\hline
 587 & 140.0 & 134.7 &  - & - & - & - \\
\hline
 1591 & - & - & 361.6 & 370.9 & - & -  \\
\hline
 899 & - & - & - & - & 94.0 & 96.7 \\
\hline
\end{tabular}
\end{table}

\begin{figure} 
\includegraphics[width=\columnwidth,keepaspectratio]{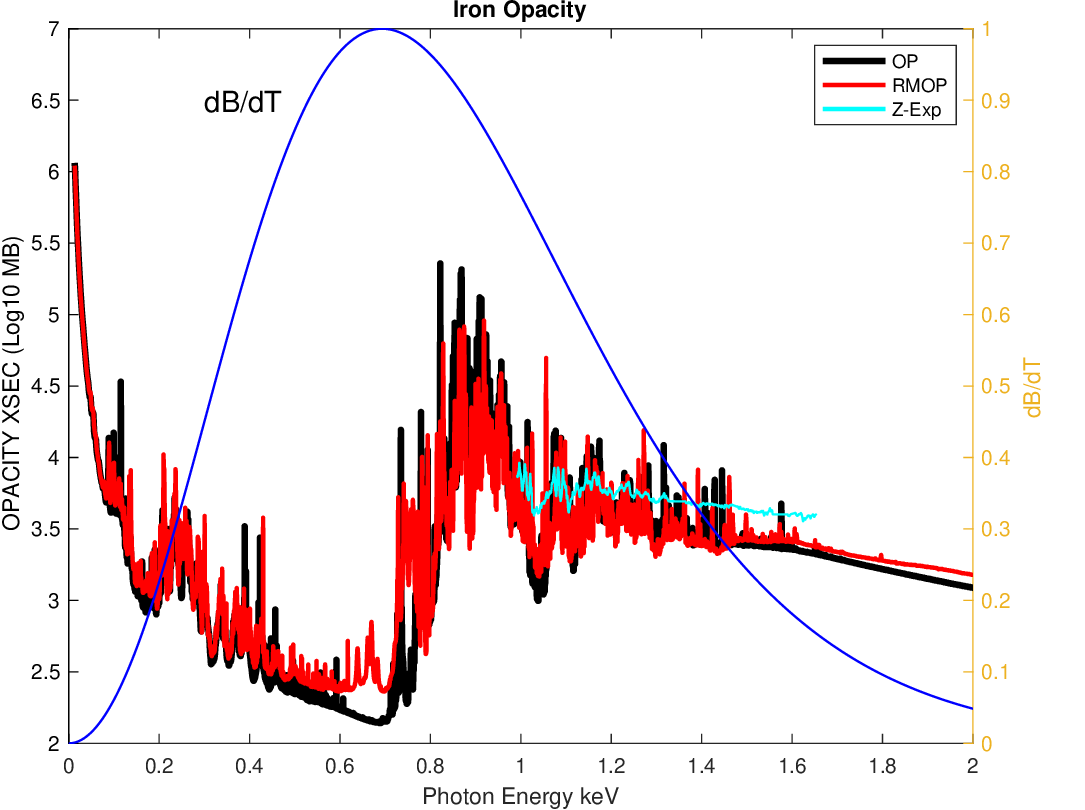}
\includegraphics[width=\columnwidth,keepaspectratio]{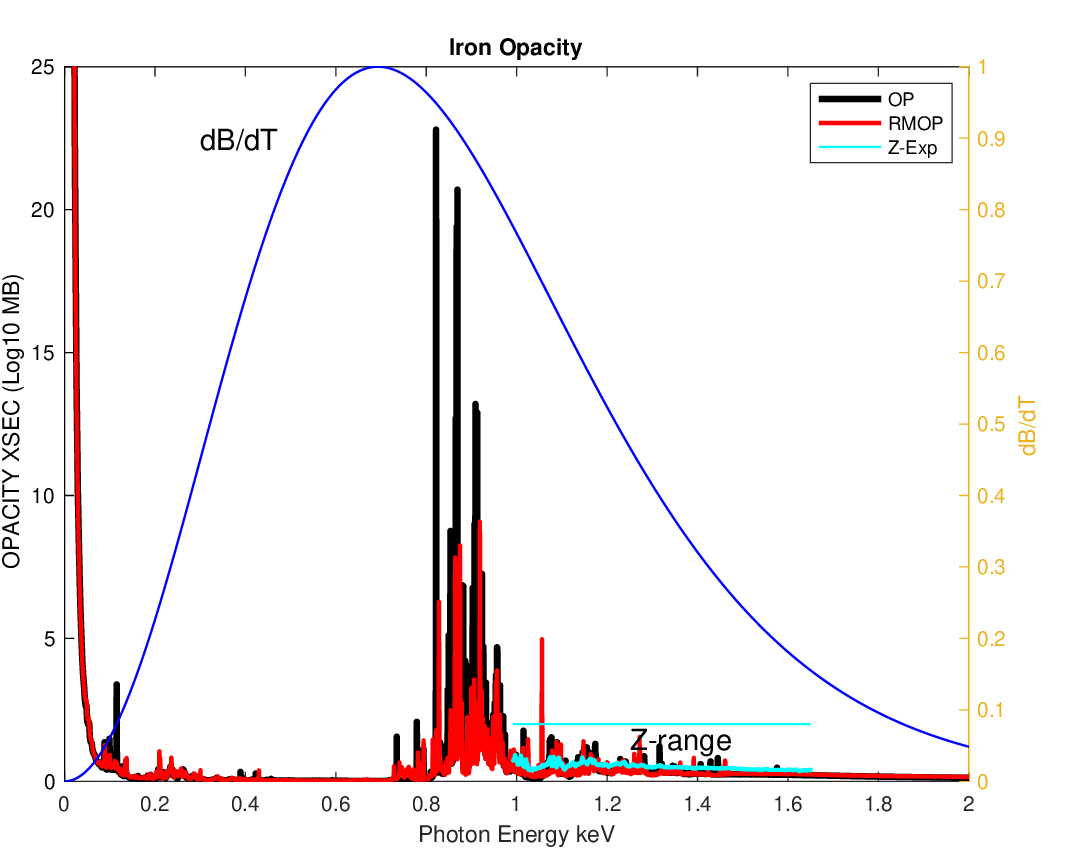}
\caption{Monochromatic opacity spectra from RMOP, OP and Sandia Z,
Log$_{10}$-scale (top) and linear values x 10$^{-4}$; the range of the
Planck function dB/dT in the Rosseland integrand is also shown. The RMOP
results demonstrate redistribution of opacity due to plasma broadening of
resonances in the bound free much more than the OP DW data. Except the
background, relative 
magnitude of experimental and theoretical data are not directly comprable
since the latter are not convolved over intrumental resolution.
 \label{fig:spec}}
\end{figure}

Fig.~\ref{fig:spec} shows a comparison of the new RMOP opacity spectrum
(red)
with OP (black). 
The Sandia Z measurements are also shown (cyan), but it should be noted
that the experimental values are convolved over instrument resolution
and the magnitudes of individual features 
are not directly compatible. In the top panel in
Fig.~\ref{fig:spec} the monochromatic opacities are plotted on a
log$_{10}$-scale, and on a linear scale in the bottom panel to better
elucidate the differences. The RMOP and OP opacity spectra differ
in detailed energy distribution and magnitude. In general, the RMOP
background is higher and the peaks lower than OP due to opacity
re-distribution, with significant enhancement around 0.7 keV.
The difference is more striking on a linear-scale in
Fig.~\ref{fig:spec} (bottom panel) around 0.9-1.0 keV, where the RMOP
peaks are lower by several factors. 

  Fig.~\ref{fig:spec} also shows that the Sandia Z
measurements span only a small energy range relative to the Planck
function derivative dB/dT that determines the Rosseland window and
therefore the RMO. But the considerable
difference between the background RMOP opacity with experiment remains 
as with the earlier OP and other works (\cite{b15,np16}. As we expect,
the background
non-resonant R-matrix photoionization cross sections are similar to 
DW results. However, the RMOP results are qualitatively in better
agreement with experimental results with shallower "windows" in
opacity than OP, for example at $E \approx 1.0$ keV (top panel) and several
other energies. Nevertheless, there seems to be a source of
background opacity in the Z experiment for iron (\cite{n19} that is
not considered in theoretical calculations. 

 It is also interesting to revisit the only available comparison 
between and OP and OPAL occupations probabilities for the simple case of
H-like C$^{5+}$ (\cite{bs03}. Table~3 gives these parameters, and also
the level populations going up to $n=6$. However, owing to the fact that
the ground state population dominates over all other levels, and Carbon
is fully ionized or H-like at given temperature-density, the RMO
remains nearly constant at 170.3 cm$^2$/g. We might expect similar
behavior for Oxygen opacity, though more detailed study is needed, and
of course for complex ions such as in this {\it Letter}.

\begin{table}
\caption{Occupation probabilities w$_n$ and level populations n-pop for H-like
C$^{5+}$ at T = 10$^6$ K, N$_e$ = 10$^{22}$ cc.
 OP opacity calculations
neglect all levels with w$_n < 10^{-3}$. Carbon is mostly fully ionized
or H-like at specified T,N$_e$: f(C$^{6+}$) = 0.431 and f(C$^{5+}$) =
0.492. RMOs are independent of EOS, $\approx$ 170 cm$^2$/g up to any level(s)
included.}
\vspace{0.1in}
\begin{tabular}{|ccccc|}
\hline
  n & w$_n$(QMHD) & w$_n$(MHD) & w$_n$ (OPAL) & Pop(n,MHD) \\ 
\hline
 1 & 1.00 & 1.00 & 1.00 & 0.438 \\ 
 2 & 0.997 & 0.983 & 0.996 & 2.42(-2) \\
 3 & 0.967 & 0.821 & 0.995 & 2.07(-2) \\
 4 & 0.705 & 0.249 & 0.995 & 8.45(-3) \\
 5 & 0.154 & 1.45(-3) & 0.914 & 6.79(-5) \\ 
 6 & 1.58(-2) & 6.0(-11) & 0.527 & 3.76(-12) \\
\hline
\end{tabular}
\end{table}

\section{Conclusion}
 Whereas improved opacities may now be computed with high precision atomic data 
using the state-of-the-art R-matrix method, the EOS remains a source of 
uncertainty. Therefore, the results presented herein should be considered tentative, pending more studies and comparison of (Q)MHD-EOS parameters with 
other equations-of-state, as well as newly improved versions (\cite{rt}.
However, preliminary RMOP results indicate considerable differences with
OP iron opacity spectrum, and by extension other existing opacity models 
based on the DW method and plasma broadening treatment of lines vs. resonances. 
While the present RMOP iron opacities are significantly higher than the OP 
owing to higher accuracy and enhanced redistribution of resonance
strengths in bound-free opacity, final results might yet depend on an improved 
MHD-EOS resolving issues outlined herein and related to pseudo bound-free
continua (\cite{dam,symp}. Although the contribution may be
relatively small around BCZ, completeness requires
R-matrix calculations for other Fe ions (in progress). 
 It is also noted that the Sandia Z experimental data are in a relatively
small energy range and therefore inconclusive as to determination of RMOs.
Although differences in background opacity with experimental data remain 
unexplained, there appears to be better agreement in detailed features.
Finally, the atomic-plasma issues described in this {\it Letter}
need to be resolved accurately in
order to obtain astrophysical opacities to solve the outstanding
solar problem.

\section*{Acknowledgments}
 I would like to thank Sultana Nahar for atomic data for Fe ions and
discussions. The computational work was 
carried out at the Ohio Supercomputer Center in Columbus 
Ohio, and the Unity cluster in the College of Arts and Sciences at the
Ohio State University. 

\section*{Data Availability}
 The data presented herein are available upon request from the author.
\label{lastpage}
\frenchspacing
\def\aa{{\it Astron. Astrophys.}\ }
\def\aasup{{\it Astron. Astrophys. Suppl. Ser.}\ }
\def\adndt{{\it Atom. data and Nucl. Data Tables.}\ }
\def\aj{{\it Astron. J.}\ }
\def\apj{{\it Astrophys. J.}\ }
\def\apjs{{\it Astrophys. J. Supp. Ser.}\ }
\def\apjl{{\it Astrophys. J. Lett.}\ }
\def\baas{{\it Bull. Amer. Astron. Soc.}\ }
\def\cpc{{\it Comput. Phys. Commun.}\ }
\def\jpb{{\it J. Phys. B}\ } 
\def\jqsrt{{\it J. Quant. Spectrosc. Radiat. Transfer}\ }
\def\mn{{\it Mon. Not. R. astr. Soc.}\ }
\def\pasp{{\it Pub. Astron. Soc. Pacific}\ }
\def\pra{{\it Phys. Rev. A}\ }
\def\pr{{\it Phys.  Rev.}\ } 
\def\prl{{\it Phys. Rev. Lett.}\ }

\bibliography{eos}

\end{document}